\documentclass[doublecol,figures]{epl2} 
\usepackage{graphicx}
\usepackage{amsmath}
\usepackage{amssymb}

\title{Generation of pure spin currents by superconducting proximity effect in quantum dots}
\shorttitle{Generation of pure spin currents by superconducting proximity effect in quantum dots} 

\author{D. Futterer\inst{1} \and M. Governale\inst{2} \and J. K\"onig\inst{1}}
\shortauthor{D. Futterer \etal}

\institute{                    
  \inst{1} Theoretische Physik, Universit\"at Duisburg-Essen and CeNIDE, 47048 Duisburg, Germany\\
  \inst{2} School of Chemical and Physical Sciences and MacDiarmid Institute for Advanced Materials and Nanotechnology, Victoria University of Wellington, PO Box 600, Wellington, New Zealand
}

 {Proximity effects; Andreev reflection; SN and SNS junctions}

\abstract{
We investigate electronic transport in a three-terminal hybrid system, composed by an interacting quantum dot tunnel coupled to one superconducting, one ferromagnetic, and one normal lead.  Despite the tendency of the charging energy to suppress the superconducting proximity effect when the quantum dot is in equilibrium, the non-equilibrium proximity effect can give rise to a large Andreev current. 
The presence of the ferromagnet can lead to a finite spin accumulation on the dot.
We find that the interplay of the Andreev current and spin accumulation can generate a pure spin current, with no associated charge transport, in the normal lead. 
This situation is realised by tuning the quantum-dot spectrum by means of a gate voltage.
}

\begin{document}
\maketitle

\section{Introduction}
Sub-gap transport through hybrid superconducting-normal interfaces is sustained by Andreev reflection \cite{andreev64,degennes:1963}.
In an Andreev-reflection process, two electrons with opposite spins enter the superconductor from the normal region forming a Cooper pair. In the reverse process, 
a Cooper pair breaks in the superconductor and its constituents, two electrons with opposite spins,  enter the normal region. 
In quantum dots Andreev reflection has been studied theoretically including Coulomb interaction and non-equilibrium \cite{theory-s-dot.fazio, theory-s-dot.kang, theory-s-dot.schwab,theory-s-dot.clerk,theory-s-dot.cuevas,theory-s-dot.pala}. 
Andreev reflection in quantum dots couples the electron and hole dynamics and it leads to the formation of  Andreev bound states. 
In the limit of infinite superconducting gap, when quasi-particle transport is inaccessible,  the coupling between quantum dot and superconductor can be treated exactly \cite{effective-hamiltonian.rozhkov,effective-hamiltonian.tanaka,effective-hamiltonian.karrasch,effective-hamiltonian.meng, governale08, futterer09}. 
Recently, several experiments have measured electronic transport through quantum dots coupled to superconducting leads \cite{cnt-qd.buitelaar02,cnt-qd.buitelaar03,cnt-qd.cleuziou,cnt-qd.jarillo-herrero,cnt-qd.ingerslev06,cnt-qd.grove-rasmussen,cnt-qd.ingerslev07,cnt-qd.eichler07,cnt-qd.eichler09,cnt-qd.herrmann,nanowire-qd.vandam,nanowire-qd.sand-jespersen,nanowire-qd.hofstetter,deacon2010,deacon2010prb}.  

Hybrid systems containing ferromagnetic materials are of interest for the possibilities they open up to generate and control spin currents. The current between two tunnel-coupled ferromagnets depends strongly on the relative magnetization direction of the two ferromagnets. For parallel alignments the tunneling current is enhanced, while for antiparallel alignments the current is suppressed. This effect is known as the tunneling magneto resistance (TMR) \cite{julliere75}.\\
In quantum dots connected to ferromagnets,  spin accumulation in the dot plays an important role.  Theoretically, for such systems complex transport properties as negative differential conductance \cite{negative-diff-conductance.bulka,negative-diff-conductance.rudzinski} and spin precession \cite{spin-precession.braig,spin-precession.braun} have been predicted. Experimentally, quantum dots realized in carbon nanotubes have already been coupled to ferromagnets \cite{ferro-cnt.sahoo,ferro-cnt.hauptmann,ferro-cnt.merchant}.\\
Cooper pairs in a BCS superconductor are made up of a spin up and spin down electron in a singlet state. Hence the combination of superconducting and ferromagnetic materials in a nano-structure is expected to give rise to rich spin physics and to non-local effects due to the entanglement intrinsic to the Cooper pairs. A good example is crossed Andreev reflection in three terminal setups consisting of a superconductor and two ferromagnets \cite{car.falci,car.lesovik,car.sanchez,car.melin,car.kalenkov}.
 
\begin{figure}
\onefigure[width=0.8\columnwidth]{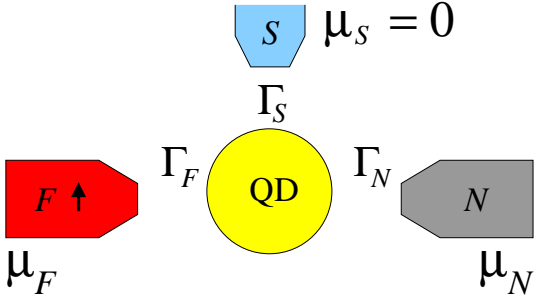}
\caption{(color online)
Setup: a quantum dot is tunnel coupled to a superconducting, a ferromagnetic, and a normal lead. 
\label{setup}}
\end{figure}

Recently, S. Das \textit{et al.} have proposed a scheme for pure spin currents in a three-terminal setup, where a ferromagnet, a superconductor, and a normal conductor are connected via quantum wires \cite{das08}.
In the present work, we consider a similar setup but involving a quantum dot.
In particular, we consider an interacting, single-level quantum dot tunnel coupled to a superconducting, a ferromagnetic, and a normal lead, see Fig. \ref{setup}. We exploit the tunability of the quantum-dot spectrum and non-equilibrium to generate a pure spin current in the normal lead,  i.e. a spin current with no charge current.

\section{Model}
The Hamiltonian of the three-terminal system under investigation can be written as $H=H_{\rm{dot}}+\sum_{\eta}(H_{\eta}+H_{\rm{tunn},\eta})$,  
where the different leads are identified by the label  $\eta= N,F,S$ corresponding  to  normal, ferromagnetic and superconducting, respectively. 
The leads are described by the Hamiltonians
\begin{equation}
H_{\eta}=\sum_{k \sigma} \epsilon_{\eta k\sigma}
c_{\eta k \sigma}^\dagger c_{\eta k \sigma}- \delta_{\eta, S} \sum_{k}\left(\Delta  c_{\eta -k \downarrow} 
c_{\eta k \uparrow}+\text{H.c.}\right), 
\end{equation}
where the single-particle energies $\epsilon_{\eta k\sigma}$ are spin dependent only for the ferromagnet, and the BCS pair-potential $\Delta$ is nonvanishing only for the superconductor. The  annihilation and creation operators for electrons in lead $\eta$ are $c_{\eta k \sigma}$ and $c_{\eta k \sigma}^\dagger$, respectively. Superconductivity is described by the  mean-field BCS Hamiltonian with a pair potential $\Delta$, which  has been chosen to be real.  The chemical potential of the superconductor is taken as reference for the energies, i.e. 
$\mu_{S}=0$.

The single-level quantum dot is described by the Anderson-impurity model:
\begin{equation} 
H_{\rm{dot}}=\sum_{\sigma} \epsilon d_{\sigma}^{\dagger} d_{\sigma} + U n_{\uparrow} 
n_{\downarrow},
\end{equation}
where $\epsilon$ is  the energy of the spin-degenerate single-particle level, and $U$ the on-site Coulomb repulsion. 
The operators $d_{\sigma}^{(\dagger)}$ annihilate (create) a dot electron with spin $\sigma$, and $n_{\sigma}=d_{\sigma}^{\dagger} d_{\sigma} $ is the corresponding number operator.
Electron tunneling between dot and leads is modeled by the   
tunneling Hamiltonians
\begin{equation}
  H_{\rm{tunn},\eta}= 
  V_{\eta} \sum_{k \sigma} \left( c_{\eta k \sigma}^\dagger d_\sigma +
  {\rm H.c.} \right), 
\label{htun}
\end{equation}
where for simplicity the tunnel-matrix 
elements $V_{\eta}$ are assumed  to be independent both of spin and momentum.
The strength of the tunnel coupling is captured by the spin-resolved tunnel-coupling strengths defined as
$\Gamma_{\eta\sigma}=2 \pi |V_{\eta}|^2 \sum_k \delta (\omega - \epsilon_{\eta k\sigma})=2\pi |V_\eta|^2 \rho_{\eta\sigma}$,  
with $\rho_{\eta\sigma}$ being  the density of states of electrons with spin $\sigma$ in lead $\eta$, which we assume to be constant in the energy window relevant for transport.
The mean level broadening is then defined as $\Gamma_\eta=\frac{1}{2}\sum_{\sigma}\Gamma_{\eta\sigma}$.
The spin asymmetry of the density of states in the ferromagnet at the Fermi energy is characterized by the spin polarization 
$p=(\rho_{F\uparrow}-\rho_{F\downarrow})/(\rho_{F\uparrow}+\rho_{F\downarrow})$.

\section{Method}
To obtain an effective description of the system, we integrate out the 
leads' degrees of freedom.
For voltages and temperatures smaller than the superconducting gap $\Delta$, quasiparticle excitations in the superconductor are suppressed.
In the following, we consider the limit $\Delta \rightarrow \infty$, for which they are completely inaccessible and only subgap-transport is possible from or to the superconductor.
This limit implies that $\Delta$ also exceeds the bias voltages applied to the leads as well as the charging energy $U$ of the dot (such that the energies for charge excitations in the dot lie within the superconducting energy gap).
The latter condition is experimentally challenging but not impossible, see e.g. Ref.~\cite{cnt-qd.jarillo-herrero}.
In the limit $\Delta\rightarrow \infty$, one can integrate out the coupling to the superconductor, which leads to an effective Hamiltonian \cite{effective-hamiltonian.rozhkov,effective-hamiltonian.tanaka,effective-hamiltonian.karrasch,effective-hamiltonian.meng} describing the proximized quantum dot:
\begin{equation}
\label{heff}
H_{\rm{eff}}=\sum_{\sigma} \epsilon d_{\sigma}^{\dagger} d_{\sigma} + U n_{\uparrow}  n_{\downarrow} -\frac{\Gamma_{S}}{2} d_{\uparrow}^{\dagger}d_{\downarrow}^{\dagger}-\frac{\Gamma_{S}}{2}d_{\downarrow}d_{\uparrow}.
\end{equation}
The Hilbert space is spanned by the four eigenstates 
$|\chi\rangle \in \{ | - \rangle, \left| \uparrow \right\rangle,
\left| \downarrow \right\rangle, | + \rangle \}$, with eigenenergies
$E_-=\frac{\delta}{2}-\epsilon_{A}$, $E_\uparrow=E_\downarrow=\epsilon$, and $E_{+}=\frac{\delta}{2}+\epsilon_{A}$, where $\left\vert\uparrow \right\rangle$ and $\left\vert \downarrow \right\rangle$ correspond to a singly occupied dot and 
\begin{eqnarray*}
\vert - \rangle &= & \frac{1}{\sqrt{2}} \left[ \sqrt{1+\frac{\delta}{2\epsilon_{A}}} \vert 0 \rangle + \sqrt{1-\frac{\delta}{2\epsilon_{A}}} \vert d \rangle \right]\\
\vert + \rangle &= & \frac{1}{\sqrt{2}} \left[  \sqrt{1-\frac{\delta}{2\epsilon_{A}}} \vert 0 \rangle - \sqrt{1+\frac{\delta}{2\epsilon_{A}}} \vert d \rangle \right] ,
\end{eqnarray*}
 are superpositions of an empty, $|0\rangle$, and doubly occupied dot, $|d\rangle=d_{\uparrow}^{\dagger}d_{\downarrow}^{\dagger}|0\rangle$ .
Here, $\delta\equiv 2\epsilon+U$ is the detuning and $\epsilon_{A}$ is given by $\epsilon_{A} \equiv \sqrt{(\epsilon + \frac{U}{2})^{2} + \frac{\Gamma_{S}^{2}}{4}}$.\\
Due to tunnel coupling to a superconductor the particle and hole sector of the Hilbert space are mixed,
leading to four Andreev bound-state energies:
\begin{equation}
\label{ABS}
E_{A, \gamma', \gamma}= \gamma' \frac{U}{2} + \gamma \epsilon_{A},
\end{equation}
with $\gamma', \gamma = \pm$. The Andreev bound-state energies are the excitation energies of the proximized dot for transitions between the singly occupied states $\left\vert \uparrow \right\rangle$ or $\left\vert \downarrow \right\rangle$ and the superposition states $\vert - \rangle$ or $\vert + \rangle$.
We consider the regime $|\delta| < \sqrt{U^2 - \Gamma_{S}^2}$, for which the inequality 
$E_{A,-,-} < E_{A,-,+} < 0 < E_{A,+,-} < E_{A,+,+}$ holds.\\

The dot is described by its reduced density matrix $\rho_{\rm{red}}$, whose matrix elements 
are $P_{\chi_2}^{\chi_1} \equiv\langle \chi_1|\rho_{\rm{red}}
|\chi_2\rangle$. 
We consider only the case of weak coupling to the non-superconducting leads,  $\Gamma_{N} ,\Gamma_{F}   \lesssim  k_{B}T$ and we compute results up to first order in these couplings.  We also assume that $\Gamma_{N} ,\Gamma_{F} \ll \epsilon_{A}$. When these conditions are fulfilled, we need to retain only the diagonal elements of the density matrix  in the basis of the eigenstates of eq.~(\ref{heff}). The diagonal elements are the occupation probabilities of the corresponding states and we denote them by $P_{\chi} \equiv P_{\chi}^{\chi}$. 

In the stationary limit, the occupation probabilities obey  the  master equation 
\begin{equation}
\sum_{\chi\ne \chi'} \left( W_{\chi \chi'}P_{\chi'}-  W_{\chi' \chi}P_{\chi}\right)=0\, ,
\end{equation}
where $ W_{\chi'' \chi'} $ is the transition rate from state $\chi'$ to state $\chi''$, which can be computed
by means of Fermi's golden rule. 
The current of electrons with spin $\sigma$ out of lead $\eta$ can be expressed as 
\begin{equation}
  J_{\eta \sigma} = \frac{1}{\hbar} \sum_{\chi\ne \chi'} 
  W_{\chi' \chi}^{ \eta \sigma} 
  P_{\chi},
\label{current}
\end{equation}
where $W_{\chi'\chi}^{ \eta \sigma} \equiv
\sum_s s W_{\chi' \chi}^{\eta \sigma}$, and 
$W_{\chi' \chi}^{\eta \sigma}$ is the sum of all rates that
describe transitions from $\chi$ to $\chi'$ in which in total $s$ electrons of spin $\sigma$ are removed from lead 
$\eta$.
The charge current and  the spin current out of lead $\eta$  are given, respectively, by
\begin{align}
J_{\eta}^{Q}  & \equiv c_{Q} \left( J_{\eta \uparrow} + J_{\eta \downarrow} \right)\\
J_{\eta}^{S}  & \equiv c_{S} \left( J_{\eta \uparrow} - J_{\eta \downarrow}\right),
\end{align}
where $c_{Q}=\rm{e}$ and $c_{S}=\hbar/2$.

\section{Results} 

We discuss the case of equal  tunnel-coupling strengths to the normal and ferromagnetic leads, $\Gamma_{N} =\Gamma_{F}  \equiv \Gamma \lesssim  k_{B}T$ , 
and focus on the regime of positive bias, $\mu_{N}>0$. 
The case $\mu_{N}<0$ is obtained from the symmetry transformation $ \mu_{N} \rightarrow -\mu_{N},\ \mu_{F} \rightarrow -\mu_{F},\ \delta \rightarrow -\delta$, and $J_{N} \rightarrow -J_{N}$.

The spin current in the normal lead 
\begin{align}
\label{jnspin}
J_{N}^{S}& = - S_{z} \Gamma_{N} \left[ \left[ f_{N}^{-}(E_{A,-,+}) + f_{N}^{+}(E_{A,+,+}) \right] \left( 1+\frac{\delta}{2\epsilon_{A}} \right) \right. \nonumber\\
&+ \left. \left[ f_{N}^{+}(E_{A,+,-}) + f_{N}^{-}(E_{A,-,-}) \right] \left( 1-\frac{\delta}{2\epsilon_{A}} \right) \right],
\end{align}
is proportional to the spin accumulation $S_{z}\equiv \left( P_{\uparrow} - P_{\downarrow} \right)/2$, where $f_{N}^{+}(\omega)=f_{N}(\omega)=[1+\exp(\frac{\omega-\mu_{N}}{k_{B}T})]^{-1}$ is the Fermi function of the normal lead and $f_{N}^{-}(\omega)=1-f_{N}(\omega)$. In eq. (\ref{jnspin}), the effect of the voltage bias applied to the ferromagnetic lead, $\mu_{F}$, is contained in the spin accumulation $S_z$. 
For small chemical potentials applied to the normal lead, $ E_{A,-,+}< \mu_{N} < E_{A,+,-}$, the spin current is exponentially suppressed. For intermediate chemical potentials of the normal lead, $E_{A,+,-} < \mu_{N} < E_{A,+,+}$, and arbitrary  $\mu_{F}$, the spin current simplifies to
\begin{align}
\label{spin-current-approx}
J_{N}^{S}  \approx - S_{z} \Gamma \left( 1-\frac{\delta}{2\epsilon_{A}} \right). 
\end{align}
In the same bias regime, the charge current is given by the expression
\begin{equation}
\label{charge-current-approx}
J_{N}^{Q} \approx  \Gamma \left[ \frac{1}{2} \left( 1-\frac{\delta}{2\epsilon_{A}} \right) P_{1} + \left( 1+ \frac{\delta}{2\epsilon_{A}} \right) P_{-} -\frac{\delta}{\epsilon_{A}} P_{+} \right],
\end{equation}
where $P_{1}=P_{\uparrow} + P_{\downarrow}$ is the probability for the dot to be singly occupied.
We note that  the charge current in intermediate bias regime can be tuned to zero by means of the dot-level position $\epsilon$. \\
On the other hand, in the large-bias regime for the normal lead, $E_{A,+,+} < \mu_{N}$, where processes with electrons entering the normal conductor are exponentially suppressed, there will always be a finite Andreev charge current flowing out of the normal lead into the superconductor, i.e., the charge current will be finite for any value of the level position.

In conclusion, a pure spin current is only possible in the intermediate-bias regime, which we will focus on in the following.

\begin{figure}
a)\\
\onefigure[width=0.8\columnwidth]{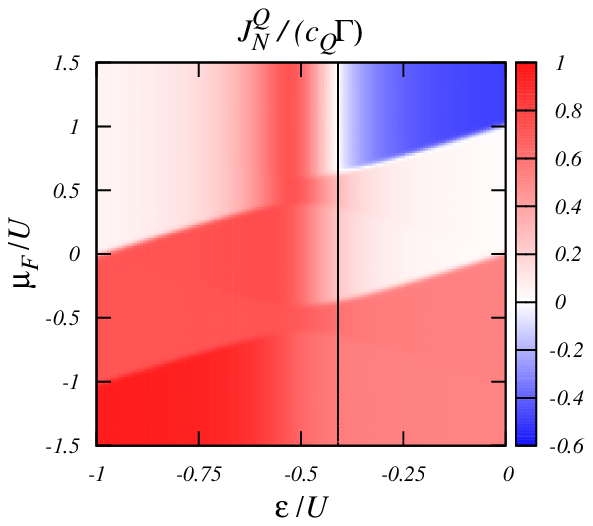}
b)\\
\onefigure[width=0.8\columnwidth]{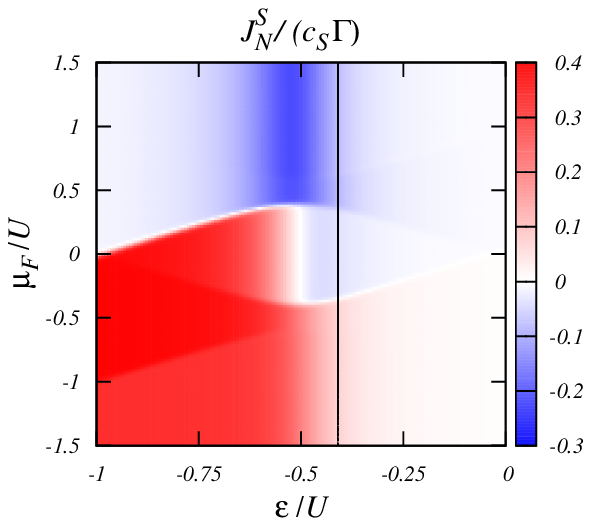}
\caption{(Color online) Density plot of a) the charge current and b) the spin current in the normal lead as a function of the level position $\epsilon$ and the chemical potential of the ferromagnetic lead $\mu_{F}$. The thin line indicates the value of the level position for which the charge current vanishes in the high-bias regime for the ferromagnetic lead. 
The other parameters are $\mu_{N}=0.5U$, $\Gamma_{S}=0.2 U$, $\Gamma=0.001 U$, $p=0.6$, and $k_{B}T=0.01 U$.  
\label{density-plots}}
\end{figure}
How to tune the system parameters to obtain a pure spin current is illustrated in Fig.  \ref{density-plots}, where the charge and spin currents are plotted as a function of the level position (tunable by the  gate voltage) and of the chemical potential of the ferromagnetic lead $\mu_{F}$ (tunable by the bias voltage). 
For the value of the level position corresponding to the thin line, the charge current vanishes for $E_{A,+,+}<\mu_{F} $, while the spin current remains finite.
The level position $\epsilon_{0}$ that tunes the charge current in the normal lead to zero, is a function of the charging energy $U$, the polarization $p$, and the tunnel-coupling strength to the superconductor $\Gamma_{S}$:
\begin{align}
\epsilon_{0} &=- \frac{U}{2}\nonumber \\
& +\frac{\Gamma_{S}}{16}  \sqrt{\frac{-p^{4}+27p^{2}-18+\left(9- p^{2} \right) \sqrt{p^{4}-36p^{2}+36}}{1-p^{2}}}.
\label{EpsilonNull}
\end{align}
The equation is valid for the normal lead being in the intermediate-bias regime, $E_{A,-,-}, E_{A,-,+}, E_{A,+,-}<\mu_{N}<E_{A,+,+}$, and the ferromagnetic lead being in the large-bias regime, $E_{A,+,+}<\mu_{F}$. 
For its derivation we have assumed  that all occurring Fermi functions can be approximated to either zero or one, i.e. that $|\mu_{N}- E_{A,\gamma',\gamma}|,\  |\mu_{F}- E_{A,\gamma',\gamma}|\gg k_{B}T $ for all possible values of $\gamma$ and $\gamma'$.

Increasing the polarization $p$ increases the spin accumulation on the dot and reduces the proximization of the dot by the superconductor. Thus the probability for the dot to be singly occupied, $P_{1}$, is increased, while the probability for the dot to be in state $|+\rangle$, $P_{+}$, is reduced. In order to keep the charge current in the normal lead tuned to zero for increased polarizations the prefactor of $P_{1}$, i.e. $(1-\delta/2\epsilon_{A})$, in eq. (\ref{charge-current-approx}) has to be reduced by increasing $\epsilon=\epsilon_{0}$.

For moderate values of the polarization $p$ the monotonically increasing function $\epsilon_{0}(p)$ grows slowly. Only for large polarizations $\epsilon_{0}(p)$ grows fast. 

When  $\epsilon_{0}$ becomes of the same order of the single-particle level spacing in the quantum dot,
the single-level model for the dot ceases to be valid. However, for the system under consideration this situation is not realized for realistic polarizations of the ferromagnetic lead. In fact, for $\Gamma_{S}=0.2 U$ and  $p = 0.995$ we find that the level position that tunes the charge current in the normal lead to zero only reaches $\epsilon_{0}\approx0$.

The pure spin current for $\epsilon=\epsilon_{0}$ is independent of the tunnel-coupling strength to the superconductor $\Gamma_{S}$ 
(though a finite $\Gamma_{S}$ is required to establish the spin current and the Andreev current tuning the charge current to zero).

\begin{figure}
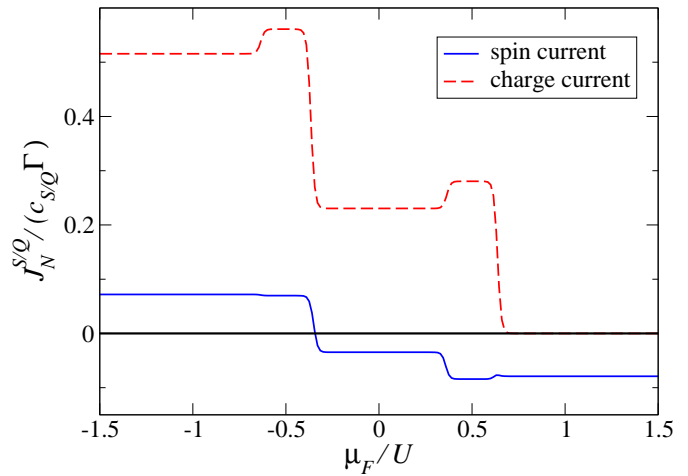

\onefigure[width=\columnwidth]{fig3.eps}
\caption{ 
(color online)
Spin current $J_{N}^{S}$ and charge current $J_{N}^{Q}$ as a function of the chemical potential of the ferromagnetic lead $\mu_{F}$ for $\epsilon=\epsilon_{0}, \mu_{N}=0.5 U,\Gamma_{S}=0.2U,\Gamma=0.001U, p=0.6,$ and $k_{B}T=0.01U$.
\label{spin-current-exp}}
\end{figure}
\begin{figure}
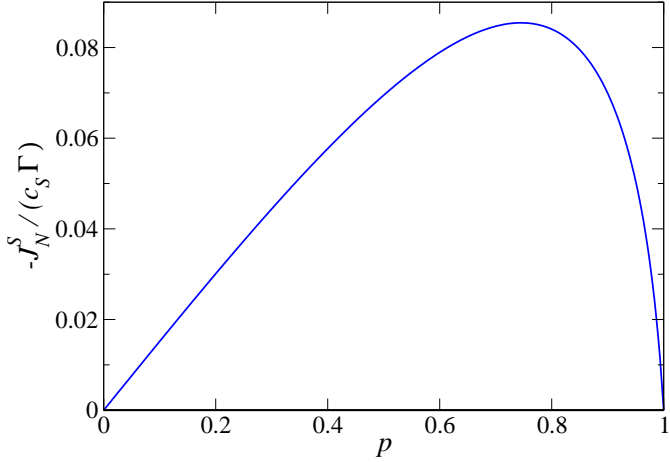

\onefigure[width=\columnwidth]{fig4.eps}
\caption{(color online) Spin current in the normal lead as a function of the polarization of the ferromagnet $p$ for $\epsilon=\epsilon_{0}, \mu_{N}=0.5U, \Gamma_{S}=0.2 U, \Gamma=0.001U$, and $k_{B}T=0.01U$.
\label{spin-current-dependence}}
\end{figure}

Figure \ref{spin-current-exp} shows the spin and charge currents in the normal lead for $\epsilon=\epsilon_0$, as a function of the chemical potential of the ferromagnetic lead $\mu_{F}$. 
The dependence of the pure spin current on the polarization of the ferromagnet is shown in Fig. \ref{spin-current-dependence}. With increasing polarization the spin current increases until a maximum is reached. For stronger polarizations, the charge current flowing out of the ferromagnet is reduced due to the spin accumulation, leading to reduction of the spin current in the normal lead.

The occurrence of a pure spin current can be understood examining the transport processes between dot and ferromagnetic / normal lead. Since the ferromagnetic and normal leads are only weakly coupled to the quantum dot, corresponding currents are carried by sequential electron tunneling.\\
\begin{figure}
\onefigure[bb=50 50 420 750,height=\columnwidth,angle={270}]{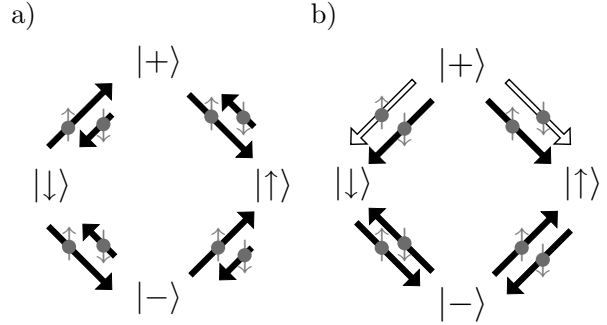}
\caption{Possible tunneling processes between dot and a) ferromagnetic lead with large chemical potential $E_{A,+,+}<\mu_{F}$ and b) normal lead with intermediate chemical potential $E_{A,+,-}<\mu_{N}<E_{A,+,+}$: the arrow direction depicts transitions of dot states. With black (white) arrows an electron leaves (enters) the corresponding lead. Tunneling of minority spins (short arrows) is suppressed. 
\label{dot-transitions}}
\end{figure}
The possible transport processes between dot and ferromagnetic lead in the large bias regime, i.e. $E_{A,+,+}<\mu_{F}$, are illustrated in Figure \ref{dot-transitions} a).  
The processes involving minority spins, depicted by shorter arrows, are suppressed. In all the processes electrons are transferred from the ferromagnetic lead to the dot. 
All four dot states can be occupied. Due to the spin polarization of the ferromagnetic lead, an imbalance in the occupation of the dot states $\left| \uparrow \right>$ and $\left| \downarrow\right>$, and thus a finite spin accumulation $S_{z}$, is generated. \\
The possible transport processes between dot and normal lead in the intermediate bias regime, i.e. $E_{A,+,-}<\mu_{N}<E_{A,+,+}$, are shown in Figure \ref{dot-transitions} b). 
Because of the spin accumulation on the dot, i.e. $P_\uparrow >P_\downarrow$, transitions from state $\left|\uparrow\right>$ to state $\left| - \right>$ take place more frequently than transitions from $\left| \downarrow \right>$ to $\left|-\right>$. This leads to a finite spin current in the normal lead. 
Black arrows represent processes where an electron flows out of the normal lead, while white arrows indicate processes where electrons enter the normal lead. 
The latter processes require the dot to be in state $\left| + \right>$.  The occupation of dot state $\left| + \right>$ is provided by the ferromagnet. By controlling the gate voltage, and thus the occupation of the different dot states, the charge current in the normal lead can be tuned to zero yielding a 
pure spin current. This pure spin current is generated by the interplay of superconductivity, non-equilibrium, and spin-dependent transport: the superconducting lead proximizes the quantum dot, leading to the superposition states $\left| - \right>$ and $\left| + \right>$. The biased ferromagnetic lead first provides occupation of the dot state $\left|+\right>$, which gives the possibility to tune the charge current in the normal to zero. Second, it causes spins to accumulate on the dot inducing a spin current in the normal lead. 

\section{Conclusions}
Spin-dependent transport in a three-teminal structure consisting of a quantum dot tunnel coupled to a superconducting, a ferromagnetic and a normal lead has been investigated in the large-superconducting-gap regime, by means of a master-equation approach. 
It has been established that such a system can be used to generate a pure spin current in the normal lead, exploiting the tunability of the dot's spectrum, non-equilibrium driven by finite bias voltages, spin accumulation due to the ferromagnet, and the Andreev bound states induced by the superconductor.

\acknowledgments
We acknowledge useful discussions with A. Saha and financial support from DFG via SFB 491.

\end{document}